\documentclass[12pt]{article}
\usepackage{pic03}
\usepackage{hyperref}
\usepackage{url}
\usepackage{graphicx}

\def \met  {\,/\!\!\!\!E_{T} }

\begin{document}

\title{\bf TOP AND ELECTROWEAK PHYSICS FROM THE TEVATRON}
\author{ 
Mark C. Kruse \\
(Representing the CDF and D\O\ collaborations)   \\
{\em Duke University, Deparment of Physics, Box 90305, Durham NC 27708}}
\maketitle

%
% photograph of author
%  This is where we will insert a photograph. To see what it would look like,
%  uncomment the following lines.
%
%\begin{figure}[h]
%\begin{center}
%
% include photograph for proceeding version
%
%\includegraphics
%[height=4.5cm]{einstein.eps}
%
% insert a fixed vertical spacing instead for the ArXiv preprint
%
\vspace{4.5cm}
%
%\end{center}
%\end{figure}

\baselineskip=14.5pt
\begin{abstract}
With the advent of the Tevatron Run 2, the collider experiments, 
CDF and D\O , have reestablished the top quark and the $W$ and $Z$ signals
and have begun programs of precision measurements of top quark 
and electroweak properties. We will survey some of the first Run 2
results from these analyses, and when appropriate compare with the
Run 1 results.
Finally we will give the status of Higgs boson searches and prospects 
at the Tevatron. 

\end{abstract}
\newpage

\baselineskip=17pt

\section{Introduction}

After the success of the Fermilab Tevatron ``Run 1'' (1992 - 1995), 
operating with protons on antiprotons with a center of mass energy of 
$\sqrt{s} = 1.8\,{\rm TeV}$ and a bunch spacing of $3.6\,{\rm \mu s}$,
a shutdown ensued to upgrade the accelerator and the collider 
experiment detectors (CDF and D\O ) for ``Run 2''.

The integrated luminosity used for physics results
from Run 1 was about $100\,{\rm pb^{-1}}$, with typical instantaneous
luminosities of $1.5\times 10^{31} {\rm cm^{-2}s^{-1}}$ (with the
best instantaneous luminosity about double this).  
Physics quality data from Run 2 started to be accumulated in earnest
in the beginning of 2002, at a higher energy of $\sqrt{s} = 1.96\,{\rm TeV}$,
and operating with a bunch spacing of 
$396\,{\rm ns}$.  The integrated luminosity goal for Run 2 has
been reduced due to difficulties with accelerator performance, however,
the aim now is accumulate between $5\,{\rm fb^{-1}}$ and $9\,{\rm fb^{-1}}$
by late this decade, which might still allow the Tevatron sufficient 
sensitivity to discover the Standard Model (SM) Higgs boson 
(or a ``SM-like'' Higgs boson),
but clearly now with a reduced chance. 
The current status of the Tevatron is given in more detail elsewhere in
these proceedings~\cite{farukh}.

\begin{figure}[htbp]
  \centerline{\hbox{ \hspace{0.2cm}
    \includegraphics[width=8.5cm]{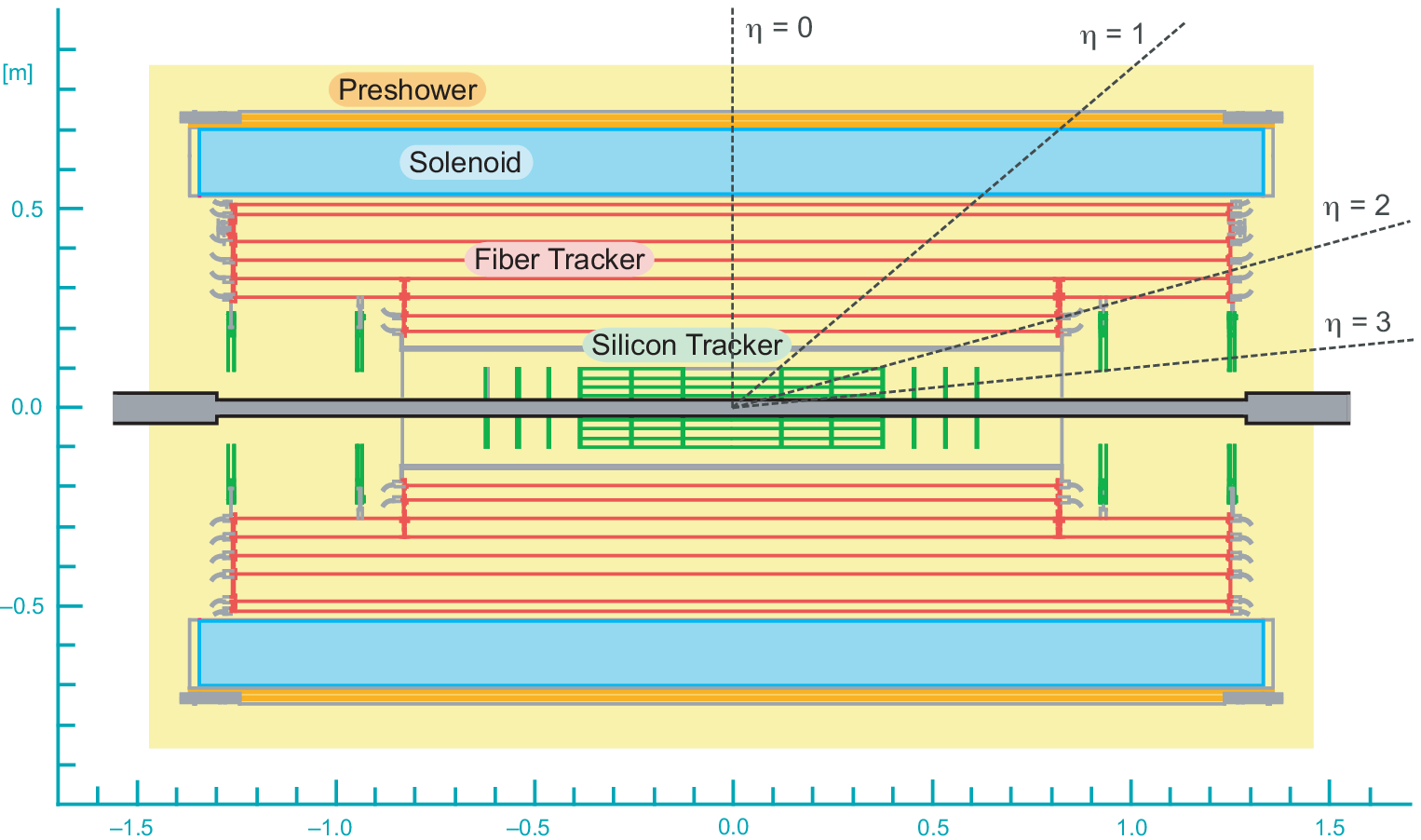}
    \hspace{0.3cm}
    \includegraphics[width=6.5cm]{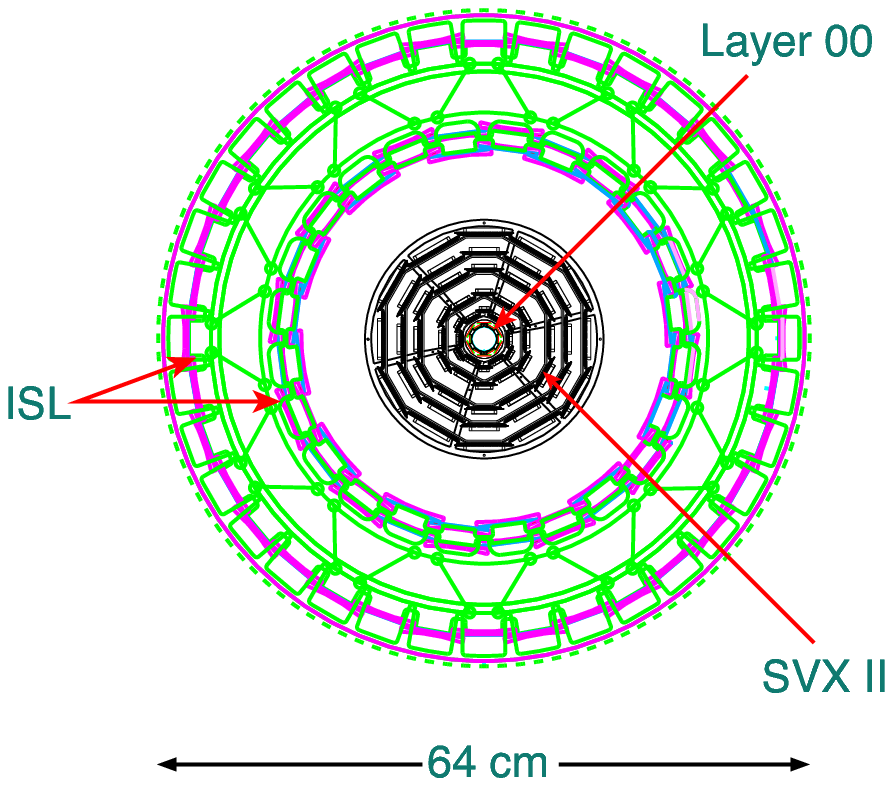}
    }
  }
 \caption{\it
      The tracking system in Run 2 for D\O\ and CDF consists of silicon
   detectors within, a Fiber tracker for D\O\  ,and a drift chamber
   for CDF. Shown here is a side view of the D\O\ tracking system (left)
   and a transverse view of the CDF silicon detectors.
    \label{fig:dets} }
\end{figure}

The CDF and D\O\ detectors are designed for general purpose use, with
a tracking system within a superconducting solenoid (1.4\,T at CDF and
2\,T at D\O\ ), calorimetry, and a muon detection system. Both detectors
underwent major upgrades for Run 2, which involved new detectors, extended
coverage of existing detector systems, and new DAQ and trigger systems.
The tracking system consists of a 
silicon system for the precise measurement of secondary 
vertices from $b$ quark decays ($c\tau = 140\,{\rm \mu m}$ which results in 
$B$ hadrons traveling on average about 
3\,mm before decaying); a central tracking system for the measurement of
particle momentum and charge, which at CDF consists of
wire chambers (COT) and a time of flight (TOF) detector (new in Run 2), and
at D\O\ consists of a Fiber Tracker. One of the important upgrades for
both experiments is the use of tracking information in the trigger system,
especially the use of silicon information to trigger on displaced vertices,
to give very high statistics $b$ and $c$ quark samples.
Figure~\ref{fig:dets} shows parts of the D\O\  and CDF tracking 
detectors.
Outside the magnetic solenoid lies the electromagnetic
and hadronic calorimeters for the measurements of energy deposition.
Electrons, photons, and jets, deposit almost all their energy in the
calorimeters. Muons travel through the calorimeters
depositing only a small fraction of their energy, and are detected by the muon
chambers (gas chambers) which surround the calorimeters and steel
absorbers.

In the following sections we will summarise a selection of past results 
using the CDF and D\O\ detectors, and discuss what we expect in the next 
several years, concentrating on the physics involving the top quark, 
some electroweak measurements, and searches for Higgs bosons. The Run 1 
results presented here come from about $100\,{\rm pb^{-1}}$ of integrated
luminosity, and the current Run 2 results from between $30\,{\rm pb^{-1}}$
and  $70\,{\rm pb^{-1}}$.

%
%\begin{figure}[htb]
%\includegraphics[width=13cm]{d0_RunII_2d-tracking_heinson.eps}
% \caption{\it
%      Example of a single figure with caption.
%    \label{exfig} }
%\end{figure}
%There are also frequent needs of placing two figures side by side
%for more efficient use of space. You can check the \LaTeX\ source
%of this template file for such an example as shown in Figure
%\ref{twofig}.

%\begin{table}
%\centering
%\caption{ \it Example of a table.
%}
%\vskip 0.1 in
%\begin{tabular}{|l|c|c|} \hline
%          &  experiment & simulation \\
%\hline
%\hline
% side-on   & $ (4.81 \pm 0.06)\%$ $E^{-\frac{1}{2}}$   &
%             $ (4.70 \pm 0.05)\%$ $E^{-\frac{1}{2}}$                    \\
% head-on   & $ (4.7 \pm 0.1)\%$ $E^{-\frac{1}{2}}$   $ + (3.4 \pm 0.6)\%$  &
%             $ (4.6 \pm 0.3)\%$ $E^{-\frac{1}{2}}$   $ + (3.8 \pm 1.3)\%$ \\
%\hline
%\end{tabular}
%\label{extab}
%\end{table}
%

\section{$W$ and $Z$ bosons}
%----------------------------

The large $W$ and $Z$ production cross-sections at the Tevatron 
(at $\sqrt{s} = 2$\,TeV;  $\sigma (p\bar{p} \rightarrow W + X) \times B(W \rightarrow \ell\nu) \sim 2.7\, {\rm nb}$,\ \ 
 $\sigma (p\bar{p} \rightarrow Z + X) \times B(Z \rightarrow \ell^+ \ell^-) 
\sim 0.26\, {\rm nb}$)
give large statistics data samples for precision measurements of vector 
boson properties. 
Only the leptonic decays are used, with the electron and muon channels
providing the cleanest signals, but with the tau channels also considered
to provide tests of lepton universality. 
The $W \rightarrow e\nu, \mu\nu$ and $Z\rightarrow e^+ e^- , \mu^+ \mu^-$
cross-section samples are also used extensively in
other analyses for the determination of energy scales, resolutions, 
lepton identification efficiencies, and checks of detector 
responses and analysis tools.

The $W \rightarrow \ell\nu$ signature is large missing transverse energy
($\met$, typically required to be $> 25\,{\rm GeV}$) and a central 
isolated lepton with
large transverse momentum ($P_T$). The backgrounds are roughly
$10\%$ for the electron and muon channels and $25\%$ for the
tau decay channel. The $Z \rightarrow \ell^+ \ell^-$ signature is
2 central high-$P_T$ leptons with opposite charge. The signal is very
clean for the electron and muon channels with backgrounds less than
$1\%$. At the time of this presentation the tau decay channel analysis
was still in progress.  

\begin{figure}[htbp]
  \centerline{\hbox{ \hspace{0.2cm}
    \includegraphics[width=8.0cm]{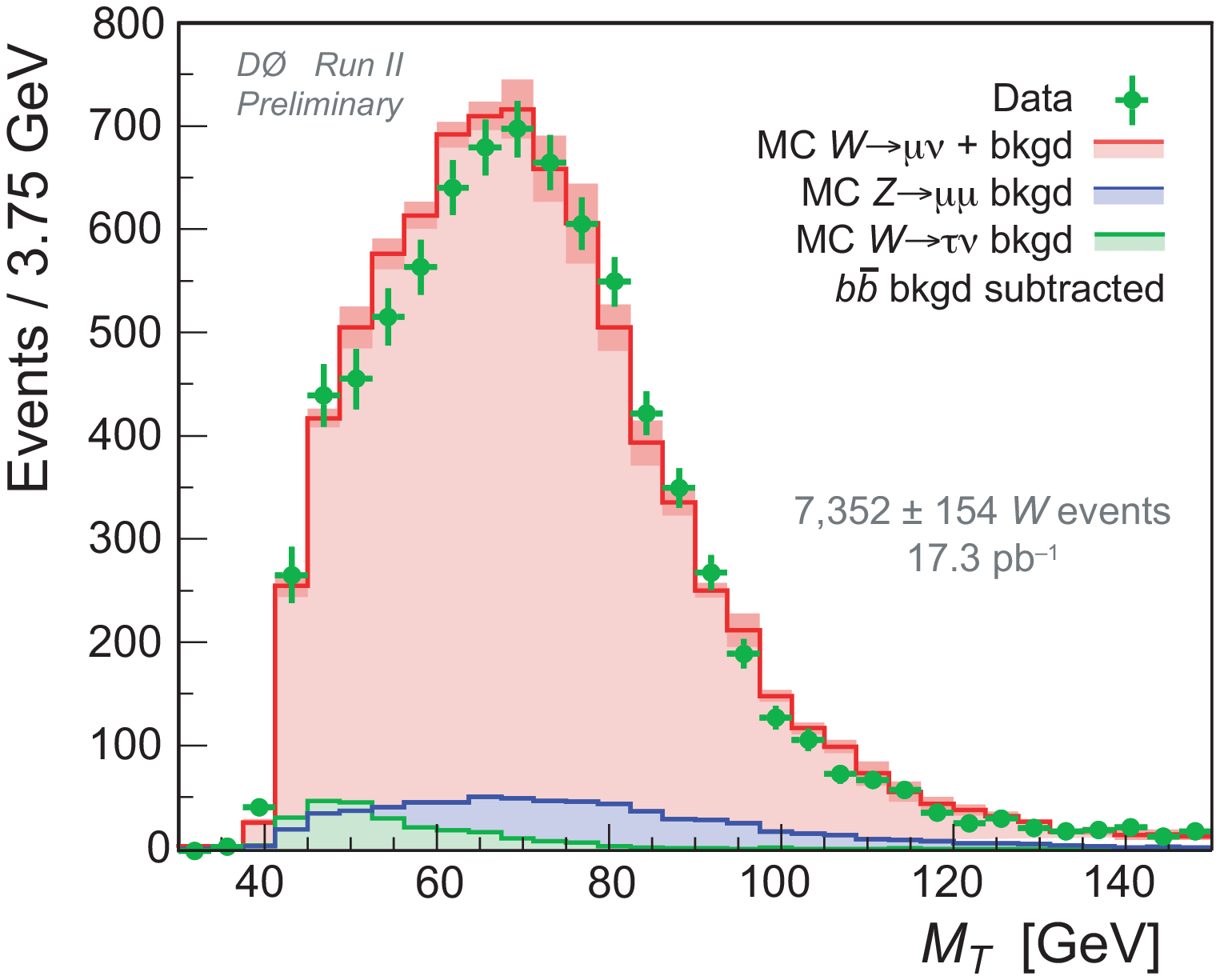}
    \hspace{0.3cm}
    \includegraphics[width=7.0cm]{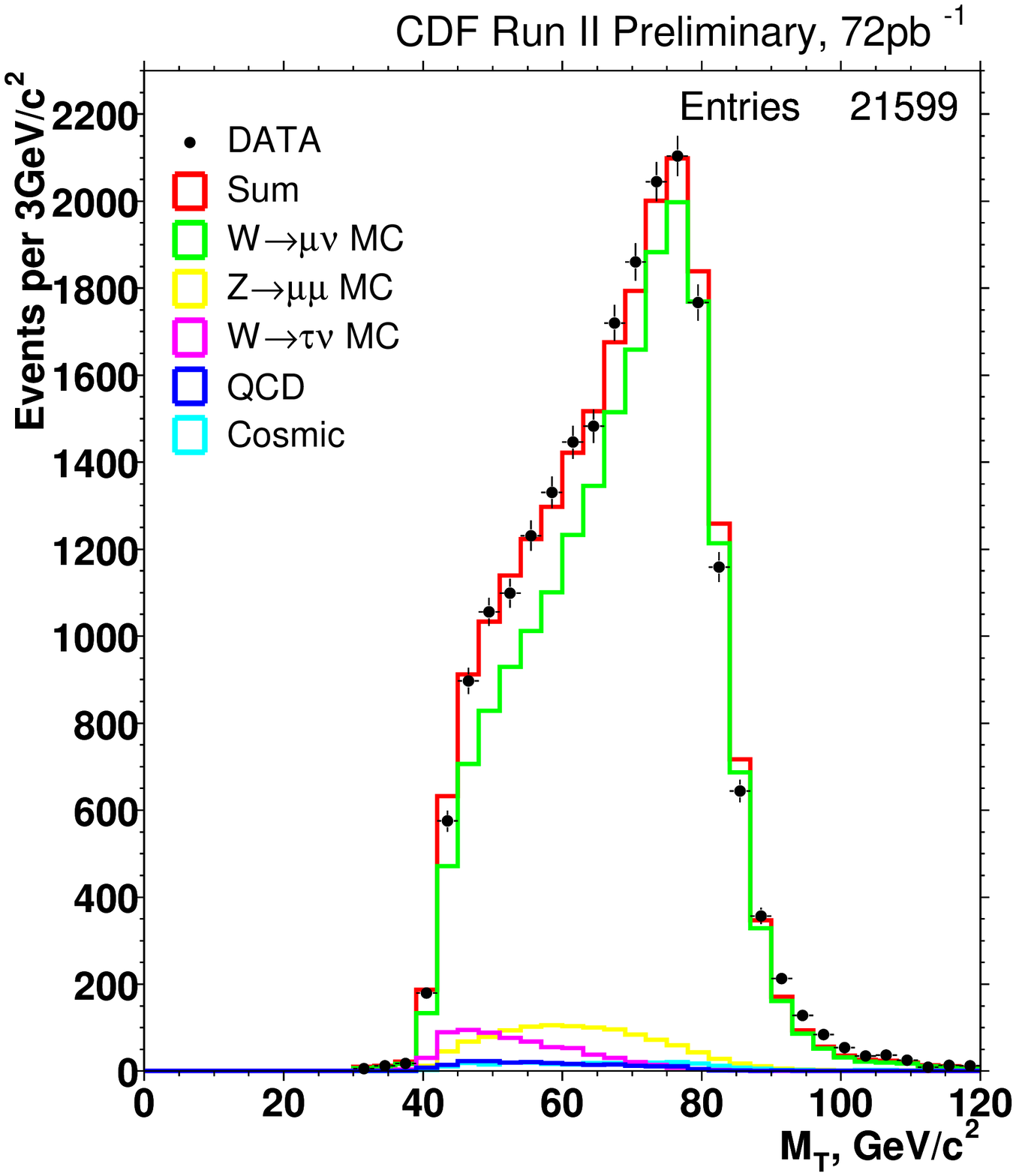}
    }
  }
 \caption{\it Transverse mass distributions of $W\rightarrow \mu\nu$ 
 candidates from D\O\ (left) and CDF (right) from the Run 2 data
 used for the cross-section measurements in Table~\ref{table:wzrun2}.
    \label{fig:wmu} }
\end{figure}

\begin{figure}[htbp]
  \centerline{\hbox{ \hspace{0.2cm}
    \includegraphics[width=7.0cm]{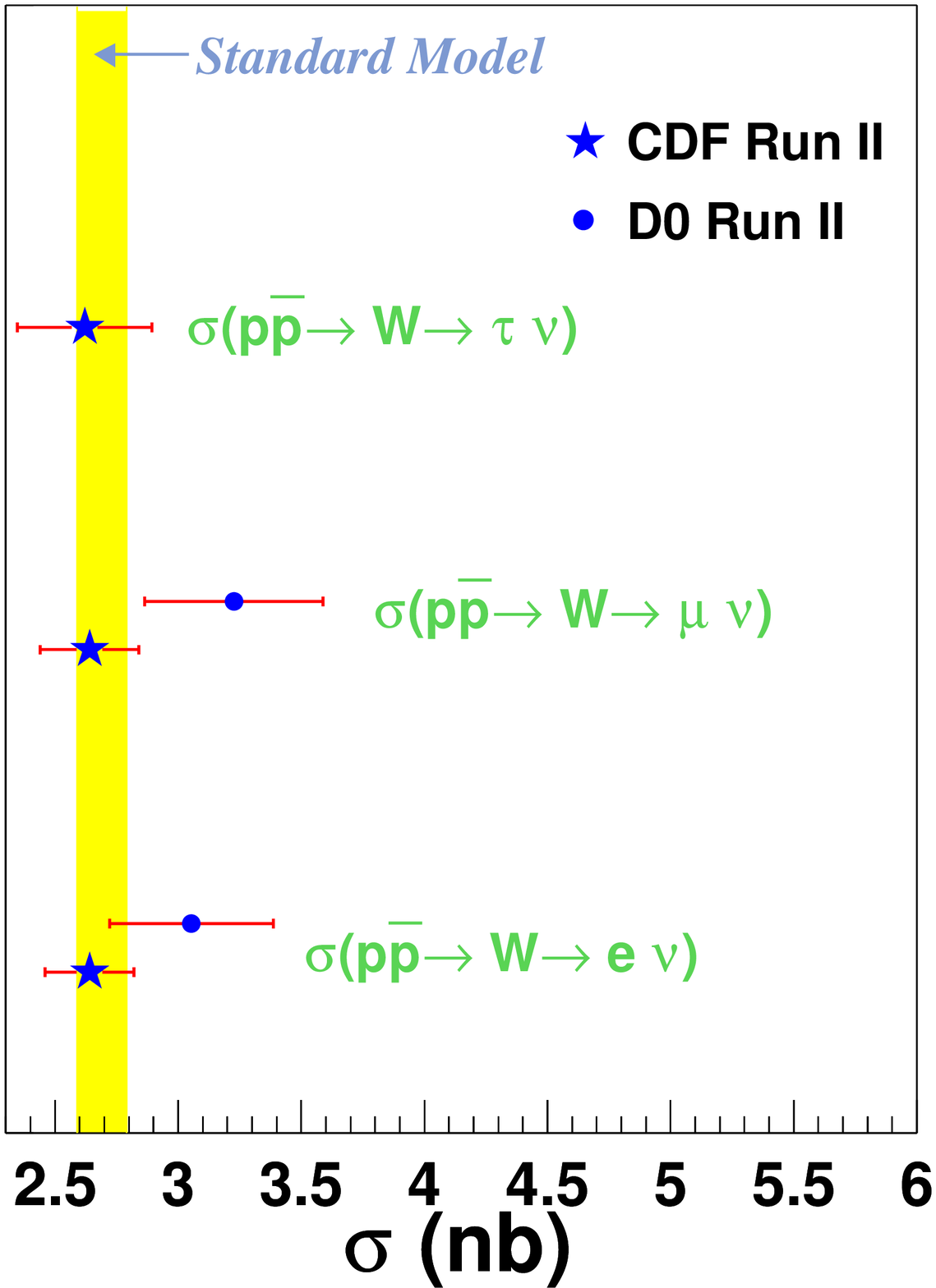}
    \hspace{0.3cm}
    \includegraphics[width=6.7cm]{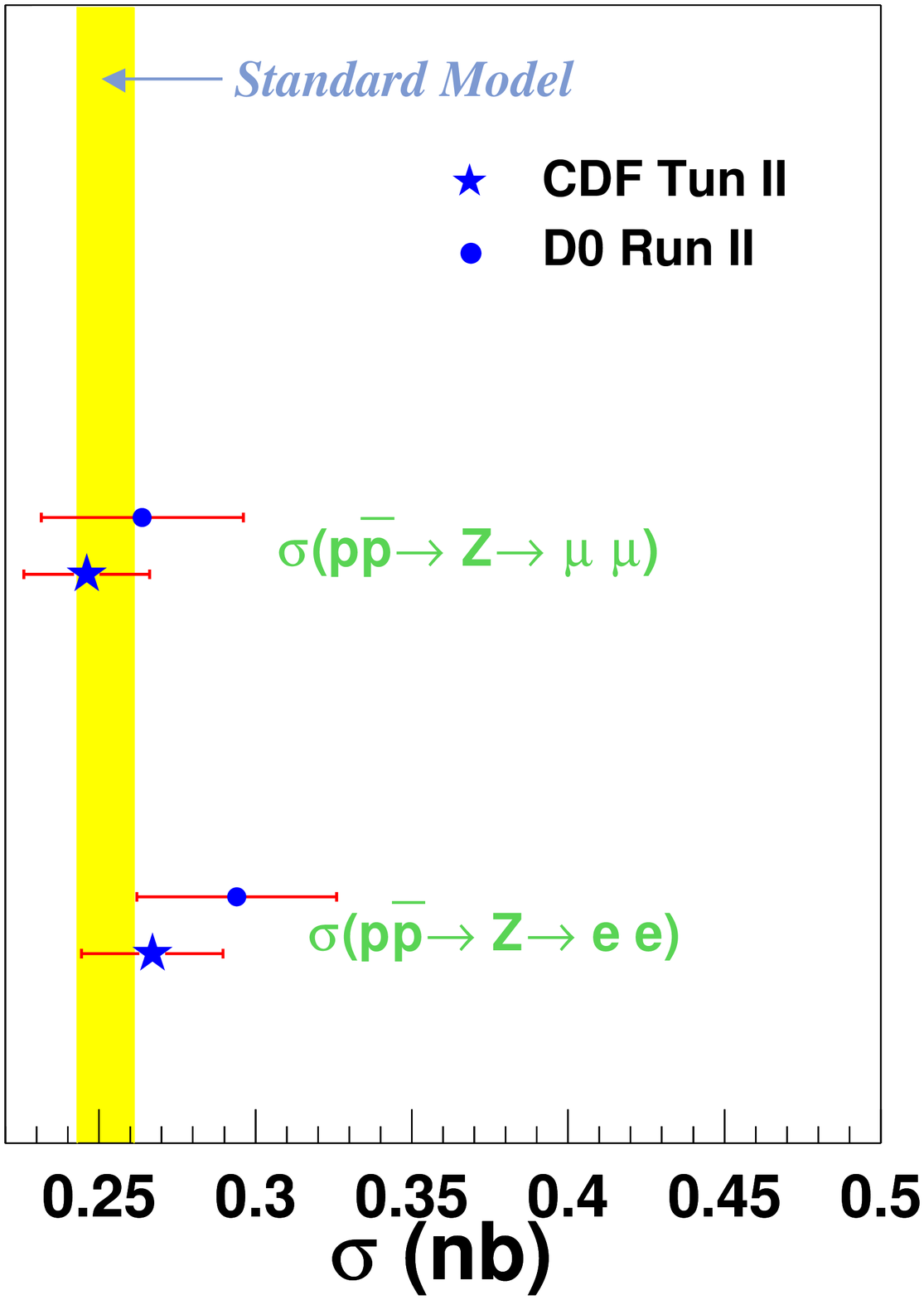}
    }
  }
 \caption{\it Comparison of the Run 2 Tevatron $W$ and $Z$ cross-section
measurements with NNLO calculations.
    \label{fig:wzcf} }
\end{figure}

The Run 2 status of the $W$ and $Z$ cross-section measurements is
summarised in Table~\ref{table:wzrun2}. Figure~\ref{fig:wmu} shows
the tranverse mass spectra used for the $W\rightarrow \mu\nu$ 
cross-section and $W$ width measurements. In Figure~\ref{fig:wzcf}
the current Run 2 $W$ and $Z$ cross-section measurements are compared
with the NNLO prediction ~\cite{wzth}.

In the near future both experiments will significantly improve the
precision of these measurements, together with various other
precision measurement not discussed here such as the $W$ mass,
$W$ width, $W$ charge asymmetry, $W$ polarization, Drell-Yan mass and 
forward-backward asymmetry.

\begin{table}
\centering
\caption{ \it Summary of Tevatron Run 2 $W$ and $Z$ production cross-section
measurements from data taken up until January 2003. The uncertainties
given are statistical, systematic, and from the luminosity measurement,
in that order.}
\vskip 0.1 in
\begin{tabular}{|l|c|c||c|c|} \hline
  & \multicolumn{2}{|c||}{D\O } &  \multicolumn{2}{|c|}{CDF} \\ \hline
 & \# events & $\sigma \times B(W\rightarrow \ell\nu)$ (nb) & 
   \# events & $\sigma \times B(W\rightarrow \ell\nu)$ (nb)  \\ \hline
$e$    & 27370 &  $3.05\pm 0.10 \pm 0.09 \pm 0.31$ 
       & 38625 &  $2.64\pm 0.01 \pm 0.09 \pm 0.16$ \\
$\mu$  & 7352  &  $3.23\pm 0.13 \pm 0.10 \pm 0.32$
     & 21599 &  $2.64\pm 0.02 \pm 0.12 \pm 0.16$ \\
$\tau$ & & 
     & 2346 &  $2.62\pm 0.07 \pm 0.21 \pm 0.16$ \\
\hline
\hline
 & \# events & $\sigma \times B(Z\rightarrow \ell\ell)$ (pb) & 
   \# events & $\sigma \times B(Z\rightarrow \ell\ell)$ (pb)  \\ \hline
$e$    & 1139  &  $294\pm 11 \pm 8 \pm 29$ 
       & 1830  &  $267\pm 6 \pm 15 \pm 16$ \\
$\mu$  & 1585  &  $264\pm 7 \pm 17 \pm 16$
       & 1631  &  $246\pm 6 \pm 12 \pm 15$ \\
\hline
\end{tabular}
\label{table:wzrun2}
\end{table}

\section{Diboson production}
%------------------------------
The cross-sections for heavy diboson production ($WW$, $WZ$, $ZZ$) are of 
the same order as for $t\bar{t}$~\cite{vvth}, but the signal is harder
to discriminate over background processes. After enough data in Run 2
these cross-sections will be measured and used to set stringent limits
on anomalous triple gauge boson couplings. The measurement and 
understanding of $WW$ and $WZ$ production will also serve as an important
precursor for associated Higgs production searches 
($q\bar{q} \rightarrow VH$ ($V = W,\,Z$)). In particular, $WZ$ production
with $Z \rightarrow b\bar{b}$ will provide a good calibration for
associated production of light Higgs with $H \rightarrow b\bar{b}$.
For heavier Higgs masses where the decay $H \rightarrow WW^\ast$
dominates, $WW$ production will be the major background, and the Higgs
searches will be natural extensions of these analyses.
Both D\O\ and CDF have started studying $WW$ production in Run 2 in the 
``dilepton'' decay channel (with a handful of candidate events), from
which both the $WW$ production cross-section measurement will come,
in addition to limits on $H \rightarrow WW^\ast$ production.

\section{Top quark physics}
%--------------------------
With Run 2 now well underway, top quark physics has moved from its discovery 
phase to precision measurements of top quark properties. With the top
quark being such an unusual object with its very large mass it may
be closely connected to new physics which precision measurements
could help unveil. 
At the Tevatron top quarks are produced predominantly in pairs via the 
strong interaction, with a cross-section at $\sqrt{s} = 1.8\,{\rm TeV}$
of about $\sigma_{t\bar{t}} = 5\,{\rm pb}$, and at $\sqrt{s} = 2.0\,{\rm TeV}$
about 7\,pb~\cite{thtop}. The top quark can also be produced singly via the
electroweak interaction with about half the $t\bar{t}$ cross-section, but 
with a final
state that is much more difficult to extract from background processes. 
The observation of single top production is one of the main goals of
the top physics program in  Run 2.
The top quark decays almost exclusively to
$Wb$, and it is the different combinations of $W$ decays that characterise
the $t\bar{t}$ decay channels. The ``dilepton'' decay channel is that
in which both $W$'s decay leptonically and has a branching ratio of
about $5\%$ (we only consider $W$ decays to $e\nu$ and $\mu\nu$ for now).
In the ``lepton + jets'' decay channel one $W$ decays leptonically and
the other hadronically (the branching ratio is about $30\%$), and in the
``all-hadronic'' decay channel both $W$'s decay hadronically (the branching
ratio is about $45\%$). 

The dilepton channel is therefore characterised
by 2 highly energetic leptons ($e$ or $\mu$ with 
$P_T > 20\,{\rm GeV}$) from the $W$ decays, a large
amount of missing energy ($\met$) from the $W$ decay neutrinos,
and 2 jets from the fragmentation of the 2 $b$ quarks. 
The main backgrounds include Drell-Yan production of $e^+ e^-$,  
$\mu^+ \mu^-$ and $\tau^+ \tau^-$ (where both taus decay to an electron or
a muon), and $WW$ production.
The lepton + jets channel is characterised by one energetic lepton,
large $\met$, and 4 jets (from the 2 $b$ quarks and the
hadronic $W$ decay). The dominant background is $W + jets$.
Lastly, the all-hadronic channel is, as its name suggests,
characterised by 6 jets only, where the dominant background is
the QCD production of jets.
Table~\ref{table:top2} summarises the expected observations from
top production at the Tevatron, and for comparison, at the LHC.

\begin{table}
\centering
\caption{ \it Summary of expected top production numbers at the
Tevatron Run 1 and Run 2, and compared with the LHC. The expected 
observed numbers in the various decay channels (the last 4 rows) use 
expected efficiencies for observing top events.}
\vskip 0.1 in
\begin{tabular}{|l|ccc|} \hline
  & Run 1 & Run 2 & LHC  \\ 
 & ($100\,{\rm pb^{-1}}$) & (per ${\rm fb^{-1}}$) & (per ${\rm 10\, fb^{-1}}$)
   \\  \hline
CM Energy (TeV)  & 1.8    & 1.96  & 14.0 \\
Peak Luminosity $(cm^{-2}s^{-1})$  & $2\times 10^{31}$ & 
      $2\times 10^{32}$ & $1 \times 10^{33}$  \\ 
 & & & \\
$\sigma(t\bar{t})\ (pb)$  & 5.0   & 7.0  & 800 \\
$\sigma$(single top)      & 2.5   & 3.4  & 320 \\
 & & & \\
Number of $(t\bar{t})$ produced & 500 &  7000 & 8,000,000  \\
Number of single top produced   & 250 &  3500 & 3,200,000 \\
 & & & \\
$N(t\bar{t}\rightarrow \ell^+ \ell^- + \met + 2\,jets)$ 
  (``dilepton'') & 4  & 80  & 50,000   \\
$N(t\bar{t}\rightarrow \ell +\geq 3j)$ ($\geq 1\ b$-tag) & 25 &700 & 400,000 \\
$N(t\bar{t}\rightarrow \ell +\geq 4j)$ ($2\ b$-tags)  & 5  & 300 & 200,000  \\
$N$(single t) ($W+2\,jets$ with $1\ b$-tag) & 3 & 70 & 60,000  \\
\hline
\end{tabular}
\label{table:top2}
\end{table}

To increase the signal to background ratio in 
the lepton + jets and all-hadronic decay channels, extensive
use is made of the silicon vertex (SVX) detectors, which can identify 
the secondary vertices from $b$ quark decays. 
In Run 1 the CDF SVX detector 
provided an efficiency for identifying at least one of the $b$ quarks
from $t\bar{t}$ decay of about $50\%$.

In addition, when $b$ quarks decay leptonically (branching ratio of about
$20\%$ for $e$ or $\mu$), one can ``tag'' the jet as coming from a $b$
quark by identifying the lepton in the jet (hereafter called an SLT tag). 
This has a much lower 
efficiency due to the difficulty in identifying these leptons and
the low branching ratio of $b$'s to leptons. For a $t\bar{t}$ event
the efficiency of tagging at least one of the $b$ jets in this way
is about $20\%$.
In Run 1 D\O\ did not have an SVX detector so made extensive 
use of this method of identifying $b$ quarks, in addition to exploiting
kinematic differences between $t\bar{t}$ and background processes, in order
to increase the signal to background ratio. 

In Run 2 both experiments
have preliminary results from using $b$ lifetime information provided
by silicon tracking. The current $b$-tagging algorithms in both experiments
are somewhat different but both have an efficiency to tag a $t\bar{t}$
event of roughly $50\%$, which will increase with more optimised algorithms
in the near future.

In Run 1 measurements of top quark properties were limited by the low
statistics available, and all errors were dominated by statistical
uncertainties. The Run 1 measurements of the top mass and cross-sections
in the various decay channels are summarised in Table~\ref{table:toprun1}.
Details of the various analyses contributing to these results, and others,
can be found in reference~\cite{top_res}.

\begin{table}
\centering
\caption{\it Summary of top quark properties as measured by the Tevatron
 from about $110\,{\rm pb^{-1}}$ of data in Run 1. The measured 
cross-sections are shown for each decay channel of $t\bar{t}$ separately,
 and for the combination of channels. The top mass shown for each 
 experiment is the result of combining the measurements in all decay
 channel.}
\begin{tabular}{|c|ccc|}\hline
Property &  CDF measurement   &  D\O\ measurement   & CDF\,+\,D\O  \\
\hline
Mass (GeV/$c^2$)  &  $176.1 \pm 6.6$   &  $172.1\pm 6.8$  & $174.3 \pm 5.1$ \\ 
$\sigma_{t\bar{t}}$ (dilepton) (pb) & $8.4^{+4.5}_{-3.5}$ & $6.4\pm 3.4$ & \\
$\sigma_{t\bar{t}}$ (lepton + jets) (pb) & $5.7^{+1.9}_{-1.5}$ & $5.2\pm 1.8$ & \\
$\sigma_{t\bar{t}}$ (all-hadronic) (pb) & $7.6^{+3.5}_{-2.7}$ & $7.1\pm 3.2$ & \\
$\sigma_{t\bar{t}}$ (combined) (pb) & $6.5^{+1.7}_{-1.4}$ & $5.9\pm 1.7$ & \\
\hline
\end{tabular}
\label{table:toprun1}
\end{table}

Preliminary Run 2 measurements of the $t\bar{t}$ production cross-section
and mass have been made, which represents the beginning of a wealth
of top physics measurements at the Tevatron in the next few years.
As the precision of the top quark measurements increases in the coming
years at the Tevatron, any possible new physics associated with the top
quark could well be uncovered.

Tables~\ref{table:dilrun2} and~\ref{table:ljrun2} summarise the first
Run 2 $t\bar{t}$ cross-section measurements at $\sqrt{s} = 1.96\,{\rm TeV}$
from the Tevatron experiments. The lepton + jets results shown in
Table~\ref{table:ljrun2} are those
that require at least one jet in the event be tagged using silicon
information. Both CDF and D\O\ employ an algorithm which reconstructs
the secondary vertex using tracks from the $b$ decay and places a
requirement on the measured distance from the primary vertex (called
an SVX tag at CDF and an SVT tag at D\O\ ). In addition D\O\ has
results using a different algorithm which requires at least 2 tracks 
in the jet with a
significant impact parameter (the perpendicular distance in the 
transverse plane from the track to the primary vertex). This is called
a ``CSIP'' tag.

In addition to the lepton + jets results shown in Table~\ref{table:ljrun2}
D\O\ also has a preliminary combined measurement using similar topological
and SLT analyses as were used in Run 1 ($W + \geq 3$ events with either a 
fourth jet or one of the 3 jets with a soft lepton tag, plus some additional
kinematic requirements). The preliminary result of combining
the topological and SLT lepton + jets analyses in about $45\,{\rm pb^{-1}}$
of Run 2 data is:
\[ \sigma_{t\bar{t}} = 
 8.5^{+4.5}_{-3.6}{\rm (stat)}^{+6.3}_{-3.5}{\rm (syst)}^{+0.8}_{-0.8}
{\rm (lum)\ pb }
\]

\begin{figure}[htbp]
%  \centerline{\hbox{ \hspace{0.2cm}
\centering
    \includegraphics[width=8.5cm]{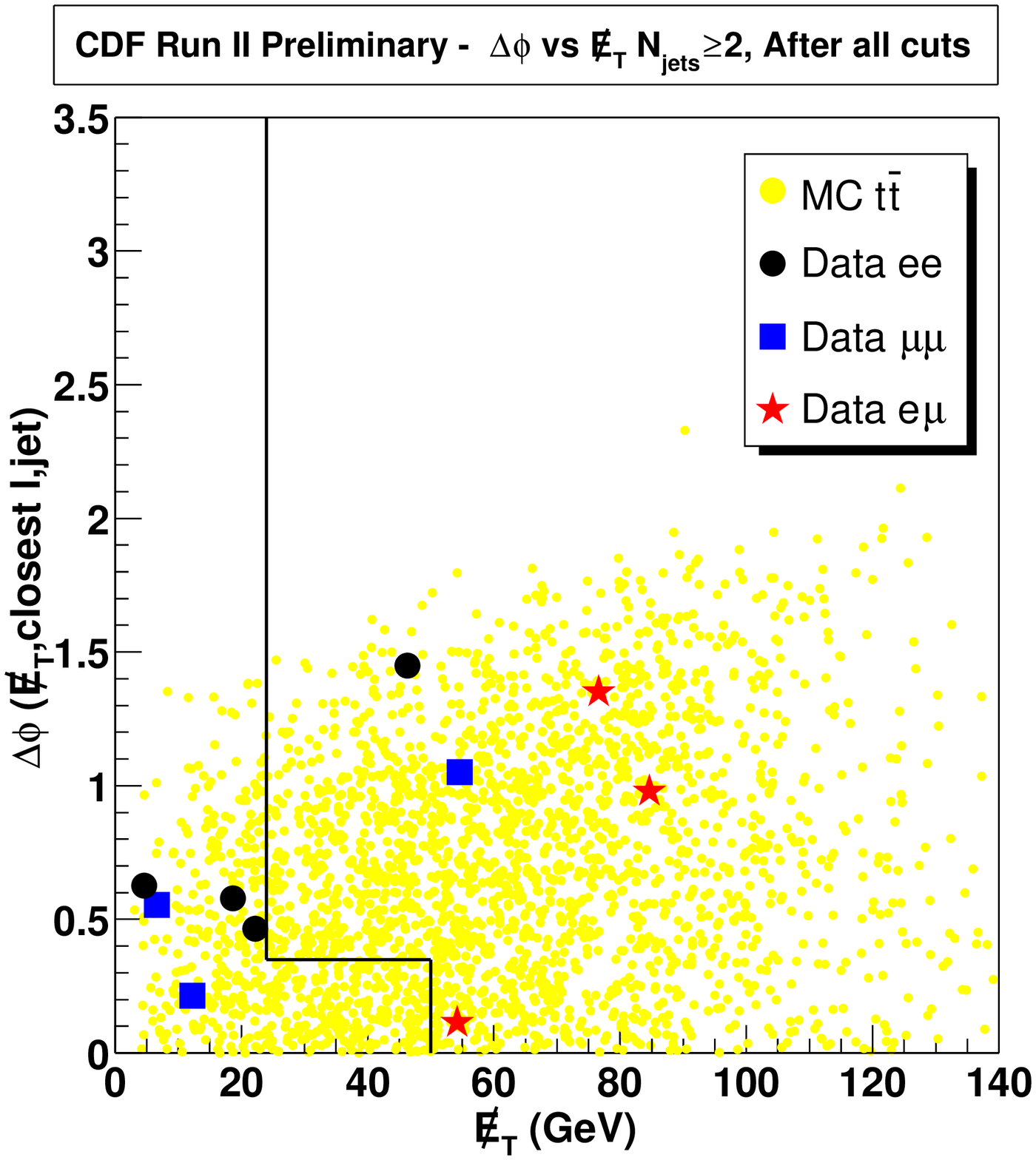}
%    \hspace{0.3cm}
%    \vspace{-0.5cm}
    \includegraphics[width=8.5cm]{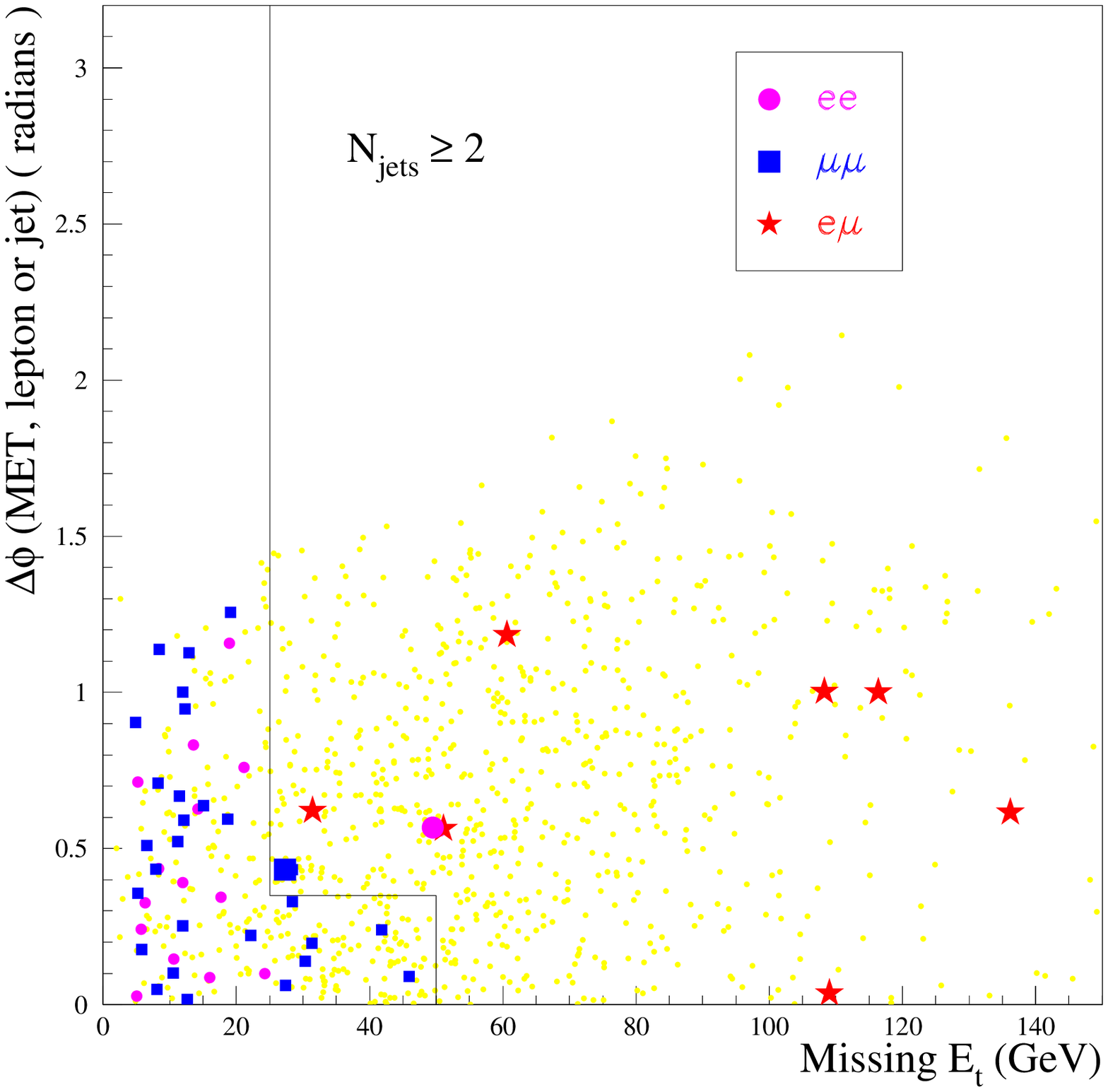}
%    }
%  }
 \vspace{-2.3cm}
     \caption{\it The CDF $t\bar{t}$ dilepton + $\geq 2$ jet events in the
  $\Delta\phi$ versus $\met$ plane, for both Run 1 (bottom) and the first 
   $72\,{\rm pb^{-1}}$ of Run 2 (top), where  $\Delta\phi$ is the azimuthal
   angle between the $\met$ and the nearest lepton or jet in the event.
   The ``L'' indicates the $\met$ and  $\Delta\phi$ requirements in the
   analysis. The Run 2 analysis has an additional cut on the total energy
   in the event, without which there is an addition 2 candidates (and so
   is a fairer comparison to the Run 1 results). In the Run 1 analysis
   9 candidates were observed with a background estimate of $2.4\pm 0.5$
   events and an expected $t\bar{t}$ contribution of $3.9$ events. 
    \label{fig:dilmet} }
\end{figure}

The Run 2 CDF $t\bar{t}$ dilepton candidates are
shown in Figure~\ref{fig:dilmet} and compared with the Run 1 results.
After $2\,{\rm fb^{-1}}$ of Run 2 data both experiments expect about
150 dilepton candidates, providing a beautiful sample of events 
in which to look for deviations from SM expectations
(expected signal to background better than 5:1 for optimised analyses).

\begin{table}
\centering
\caption{ \it Preliminary Run 2 $t\bar{t}$ production cross-section
  measurements from D\O\ and CDF in the dilepton channel. The Run 2 
integrated luminosity for these measurements is $72\,{\rm pb^{-1}}$
for CDF, and for D\O\  $48\,{\rm pb^{-1}}$ for $ee$, $33\,{\rm pb^{-1}}$ 
for $\mu\mu$, and $42\,{\rm pb^{-1}}$ for $e\mu$.
 }
\vskip 0.1 in
\begin{tabular}{|l|ccc|} \hline
D\O   &  $ee$  & $\mu\mu$ & $e\mu$ \\
\hline
Background          & $1.00\pm 0.48$ & $0.59\pm 0.30$  & $0.07\pm 0.01$  \\
Expected $t\bar{t}$ & $0.25\pm 0.02$ & $0.30\pm 0.02$  & $0.50\pm 0.01$  \\
\hline
Run 2 data & 4 & 2 & 1 \\
\hline
$\sigma_{t\bar{t}}$ (pb) & 
\multicolumn{3}{|c|}{  
$29.9^{+21.0}_{-15.7}{\rm (stat)}^{+14.1}_{-6.1}{\rm (syst)}^{+3.0}_{-3.0}{\rm (lum)}$} \\
\hline\hline
 CDF &  $ee$  & $\mu\mu$ & $e\mu$  \\ \hline
Background &  $0.10\pm 0.06$ & $0.09\pm 0.05$  & $0.10\pm 0.04$ \\
Expected $t\bar{t}$ & $0.47\pm 0.05$ & $0.59\pm 0.07$  & $1.44\pm 0.16$ \\
\hline
Run 2 data & 1 & 1 & 3 \\
\hline
$\sigma_{t\bar{t}}$ (pb) & \multicolumn{3}{|c|}{  
$13.2^{+7.3}_{-5.4}{\rm (stat)}^{+2.0}_{-0.8}{\rm (syst)}^{+0.8}_{-0.8}{\rm (lum)}$} \\
\hline
\end{tabular}
\label{table:dilrun2}
\end{table}

\begin{table}
\centering
\caption{ \it Preliminary Run 2 $t\bar{t}$ production cross-section
  measurements from D\O\ and CDF in the lepton + jets channel where
at least one jet is required to be tagged by the SVX. Results
are shown from D\O\ using 2 different secondary vertex tagging algorithms. 
The Run 2 
integrated luminosity for these measurements is $58\,{\rm pb^{-1}}$
for CDF, and for D\O\  about $45\,{\rm pb^{-1}}$. The 3 and $\geq 4$ jet 
events are used for the cross-section measurements. 
 }
\vskip 0.1 in
\begin{tabular}{|l|cc|cc|} \hline
D\O\ (CSIP) & $W + 1$ jet  & $W + 2$ jets & $W + 3$ jets & $W + \geq 4$ jets \\
\hline
Background  & 
   $30.6\pm 5.0$ & $26.4\pm 3.5$  & $8.3\pm 1.3$ & $2.5 \pm 0.7$ \\
SM Background + $t\bar{t}$ & 
   $30.6\pm 5.0$ & $27.1\pm 3.6$  & $11.1\pm 1.4$ & $6.5 \pm 1.0$ \\
\hline
Run 2 data & 34 & 27 & 13 & 6 \\
\hline
$\sigma_{t\bar{t}}$ (pb) & \multicolumn{4}{|c|}{  
$7.4^{+4.4}_{-3.6}{\rm (stat)}^{+2.1}_{-1.8}{\rm (syst)}^{+0.7}_{-0.7}{\rm (lum)}$} \\
\hline\hline
D\O\ (SVT) & $W + 1$ jet  & $W + 2$ jets & $W + 3$ jets & $W + \geq 4$ jets \\
\hline
Background  & 
   $27.0\pm 5.0$ & $24.0\pm 4.0$  & $7.5\pm 1.5$ & $2.3 \pm 0.6$ \\
SM Background + $t\bar{t}$ & 
   $27.0\pm 5.0$ & $24.6\pm 4.1$  & $10.1\pm 1.6$ & $6.0 \pm 0.9$ \\
\hline
Run 2 data & 28 & 20 & 9 & 9 \\
\hline
$\sigma_{t\bar{t}}$ (pb) & \multicolumn{4}{|c|}{  
$10.8^{+4.9}_{-4.0}{\rm (stat)}^{+2.1}_{-2.0}{\rm (syst)}^{+1.1}_{-1.1}{\rm (lum)}$} \\
\hline\hline
 CDF (SVX) & $W + 1$ jet  & $W + 2$ jets & $W + 3$ jets & $W + \geq 4$ jets \\
\hline
Background &  
   $33.8\pm 5.0$ & $16.4\pm 2.4$  & $2.9\pm 0.5$ & $0.9 \pm 0.2$ \\
SM Background + $t\bar{t}$ & 
   $34.0\pm 5.0$ & $18.7\pm 2.4$  & $7.4\pm 1.4$ & $7.6 \pm 2.0$ \\
\hline
Run 2 data & 31 & 26 & 7 & 8 \\
\hline
$\sigma_{t\bar{t}}$ (pb) & \multicolumn{4}{|c|}{  
$5.3^{+2.1}_{-1.8}{\rm (stat)}^{+1.3}_{-0.6}{\rm (syst)}^{+0.3}_{-0.3}{\rm (lum)}$} \\
\hline
\end{tabular}
\label{table:ljrun2}
\end{table}

\begin{figure}[htb]
\centering
\includegraphics[width=9.0cm]{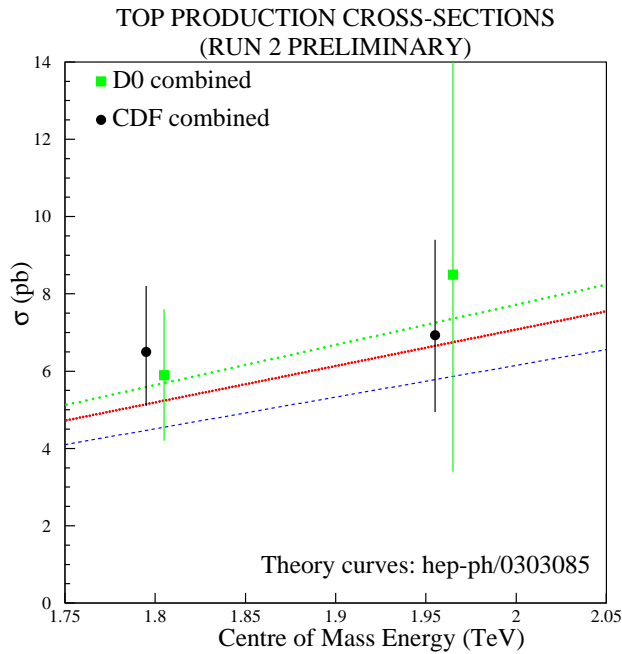}
 \caption{\it Summary of the D\O\  and CDF Run 1 $t\bar{t}$ production 
  cross-section measurements at $\sqrt{s} = 1.8\,{\rm TeV}$, together
  with the first Run 2 results, as compared with a NLO calculation.
    \label{fig:xsecvss} }
\end{figure}

As Run 2 progresses the statistical limitations of the results 
presented here will be reduced dramatically, and after a few hundred
${\rm pb^{-1}}$ of data the main limitations will be systematic. 
At the time of this writing new cross-section measurements are about
to come out using more than $100\,{\rm pb^{-1}}$, and this continual
improvement of top quark measurements will continue for the next several
years. The top cross-section measurements from Run 1, together with the
first Run 2 results are summarised in Figure~\ref{fig:xsecvss}. The Run 2 
results shown are only the combined D\O\ topological and SLT cross-section
given above, and the combination of the CDF lepton + jets and dilepton
cross-sections, which gives a result of 
$\sigma_{t\bar{t}} = 6.9^{+2.5}_{-2.0}\ {\rm pb}$.
Shown for comparison is a NLO prediction~\cite{thtop}  where the upper and
   lower curves represent doubling and halving the central scale of $m_{top}$
   and incorporate the spread from different PDF's.

Run 2 measurements of the top mass in both the dilepton and lepton + jets 
channels are also in progress. After accumulating approximately 
$1\,{\rm fb^{-1}}$ of Run 2 data, it is expected the $t\bar{t}$ production
cross-section will be measured to about $10\%$, and the top mass
to about 3\,GeV. In both cases combinations of D\O\ and CDF analyses
are planned to improve the overall Tevatron sensitivity. 

\section{Searches for the Higgs}
%-------------------------------
Within the SM mass is generated, and the electroweak symmetry is broken,
by the Higgs mechanism. All the particles of the SM have been observed,
except for the Higgs boson, so its discovery is currently one of the most
important goals in high energy physics. Precision electroweak measurements
(including the top mass) predict a light SM Higgs~\cite{ewklep} of
$88^{+53}_{-35}\,{\rm GeV}/c^2$. The expected light Higgs makes it possible
to discover at the Tevatron in the next few years, although with the 
reduced predicted Run 2 integrated luminosity this goal is now less
likely.
In Run 1 CDF and D\O\ conducted searches
for low mass Higgs bosons ($M_H < 120\,{\rm GeV}/c^2$), using its associated
production with vector bosons ($W^{\pm}$ or $Z^0$). At such masses the
Higgs decays predominantly to $b\bar{b}$. Therefore, even though the
cross-section for $gg \rightarrow H$ (0.7\,pb at $M_H = 120\,{\rm GeV}/c^2$) 
is 
much larger than the cross-sections for $WH$ and $ZH$ production 
(0.16\,pb and 0.10\,pb respectively at $M_H = 120\,{\rm GeV}/c^2$ 
and $\sqrt{s} = 2$\,TeV)~\cite{spira}, 
the Tevatron experiments are more sensitive to the latter production
mechanisms, as single Higgs production is overwhelmed by an irreducible 
QCD di-jet background. 

In Run 1 searches were conducted in the following associated SM Higgs 
decay channels:
\begin{itemize}
\item $ZH \rightarrow \ell\ell\,b\bar{b}$ (with at least 1 $b$-tagged jet): 
signal characterised by 2 highly energetic leptons ($P_T > 20$\, GeV) and 
 2 jets.  
\item $ZH \rightarrow \nu\nu\, b\bar{b}$ (with at least one $b$-tagged jet):
  signal characterised by large missing energy and 2 jets.
\item $WH \rightarrow \ell\nu\, b\bar{b}$ (with at least one $b$-tagged jet):
 signal characterised by one energetic lepton, large missing energy and
 2 jets.
\item $W/Z\,H \rightarrow qq^\prime \,b\bar{b}$ (with at least two $b$-tagged 
  jets): signal characterised by 4 jets.
\end{itemize}
 
\begin{figure}[htb]
\centering
\vspace*{+1.5cm}
\includegraphics[width=10cm]{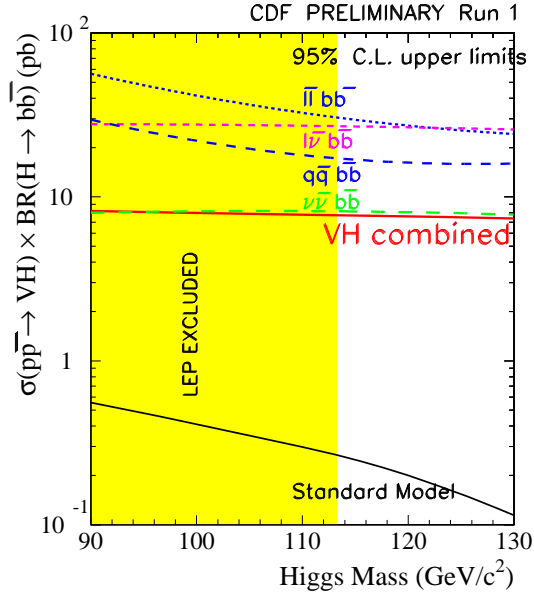}
 \caption{\it
     CDF Run 1 results from searches for 
   the SM Higgs boson.  The 95\% confidence level 
   limits for $WH + ZH$ production are shown
   as a function of Higgs mass for the different decay channels analysed.
   Also shown is the region excluded by LEP2.
    \label{fig:vh} }
\end{figure}

In the absence of any observed signal,
cross-section limits were determined by 
fitting the observed spectra to a combination of the expected di-jet spectra 
from background processes and from $WH$ and $ZH$ signal. Figure~\ref{fig:vh}
summarises the CDF 95\% confidence level (CL) cross-section limits for all 
the individual decay channels as well as the combined result, as a function 
of the Higgs mass. The limits are about 30 times higher than the SM
predictions, which sets the scale for the luminosity needed in order to
observe the SM Higgs boson in Run 2.

In Run 2 $b$-tagging and the resolution of the $b\bar{b}$ invariant mass
is critical for light Higgs searches. To this end, the success of the CDF
and D\O\ silicon vertex triggers is crucial, from which a large
$Z \rightarrow b\bar{b}$ sample will be obtained for calibrating the
tagging efficiency and mass resolution. It is also important for
CDF and D\O\ to combine results for maximal sensitivity. Run 2 sensitivity
studies indicate that with about $10\,{\rm fb^{-1}}$ of data (the current
upper limit projected for Run 2) evidence for Higgs production at the
3$\sigma$ level is possible for $M_H < 130\,{\rm GeV}/c^2$, and exclusion
at the 95\% level is possible up to a Higgs mass of 
$180\,{\rm GeV}/c^2$~\cite{run2higgs}. In Run 2 searches for $H \rightarrow
W^+ W^-$ are also in progress which will provide limits in the intermediate
Higgs mass range (assuming something isn't found).

\section{Conclusions}
%--------------------

Both the CDF and D\O\ detectors are working well in Run 2 and the understanding
of their performance is now leading to high quality physics results.
We have reported on preliminary Run 2 results in electroweak and top quark
physics, and summarised the relevant Run 1 results for comparison. With
the recent improvements in the Tevatron's performance, these Run 2 results
represent the beginning of a rich program in electroweak, top quark, and
Higgs physics in the next few years.

\section{Acknowledgements}
%-------------------------
I would like to thank the organisers of the PIC 2003 conference
for a stimulating and well organised series of presentations. 
This work 
was supported by the U.S. Department of Energy and National Science Foundation;
the Italian Istituto Nazionale di Fisica Nucleare; the Ministry of Education,
Science and Culture of Japan; the Natural Sciences and Engineering Research
Council of Canada; the National Science Council of the Republic of China;
and the A.P. Sloan Foundation.

%\section{References}
%References \cite{fitch} in bibliography must be referred to in the
%text enclosed in brackets. All references should be organized to
%provide initials and last name of the first author, publication
%title, volume (bold-face), page number, year (in brackets) of
%publication.

\end{document}